# On the Asymptotic Weight and Stopping Set Distribution of Regular LDPC Ensembles

Vishwambhar Rathi *

**Abstract**

We estimate the variance of weight and stopping set distribution of regular LDPC ensembles. Using this estimate and the second moment method we obtain bounds on the probability that a randomly chosen code from regular LDPC ensemble has its weight distribution and stopping set distribution close to respective ensemble averages. We are able to show that a large fraction of total number of codes have their weight and stopping set distribution close to the average.

## 1 Introduction

The weight distribution is an important characterization of a code. For a code $G$ of block length $n$, we define $N(G,n\omega)$ as the weight distribution function, denoting the number of codewords with normalized weight $\omega$[1]. In general $N(G,n\omega)$ is hard to compute for a specific code. In fact, even the determination of the minimum distance is NP-complete [18]. On the other hand, for some ensembles of codes it is easy to compute the *expected* weight distribution function, i.e., $\mathbb{E}[N(G,n\omega)]$. This is true for e.g. Shannon's random ensemble but also for suitably defined LDPC ensembles. A possible approach to study the weight distribution of *individual* codes is to first compute the ensemble average and then to show that most codes have a weight distribution close to this average. For LDPC codes it has been conjectured that for regular ensembles most codes have a weight distribution close to the ensemble average [2, 13].

In 1989, Sourlas showed that there is a strong connection between error-correcting codes and disordered spin models [16, 17]. To this end, let us consider the exponent $\frac{1}{n}\ln N(G,n\omega)$ and define:

$$
\begin{aligned}
W_{\text{sp}}(\omega) &:= \lim_{n\to\infty}\frac{1}{n}\mathbb{E}[\ln N(G,n\omega)],\\
W_{\text{com}}(\omega) &:= \lim_{n\to\infty}\frac{1}{n}\ln\mathbb{E}[N(G,n\omega)],
\end{aligned}
$$

where sp stands for "statistical physics", since $W_{\text{sp}}(\omega)$ can be computed by statistical physics methods and com stands for "combinatorics", as $W_{\text{com}}(\omega)$ can easily be computed by combinatorial methods. From Jensen's inequality we know that $W_{sp}(\omega) \leq W_{com}(\omega)$. It has been shown in [2, 13] that for regular LDPC ensembles $W_{sp}(\omega) = W_{com}(\omega)$ . However for irregular LDPC ensembles this is not the case [4]. The equality between $W_{com}(\omega)$ and $W_{sp}(\omega)$ for regular ensembles suggests that a randomly chosen code should have $N(G, n\omega)$ "close" to $\mathbb{E}[N(G, n\omega)]$ with high probability. In this paper we obtain an asymptotic lower bound on this probability using the second moment method by estimating the variance of $N(G, n\omega)$. However, to estimate the variance we need to verify that the solution set of a certain system of polynomial equations satisfies some properties (see Lemma 3.4 for details). Assuming that these properties are satisfied, we show that for a regular LDPC ensemble with left degree $\mathtt{l}$ and right degree $\mathtt{r}$, any $\epsilon > 0$ and for all $\omega$ such that $W_{com}(\omega)$ is positive,

$$\lim_{n\to\infty} \mathrm{P}\left(1 - \epsilon \leq \frac{N(G, n\omega)}{\mathbb{E}[N(G, n\omega)]} \leq 1 + \epsilon\right) \geq 1 - \frac{\delta(\omega, \mathtt{l}, \mathtt{r})}{\epsilon^2}, \tag{1}$$

where $\delta(\omega, \mathtt{l}, \mathtt{r})$ is a function of $\omega$ and can be evaluated by solving a polynomial equation. If $W_{com}(\omega) \leq 0$, then $\mathbb{E}[N(G, n\omega)] = o(1)$ (Lemma 3.2) and by Markov's inequality we have

$$\lim_{n\to\infty} \mathrm{P}\left(N(G, n\omega) = 0\right) = 1.$$

*EPFL (Lausanne), CH-1015, email:`vishwambhar.rathi@epfl.ch`

[1]Here and in what follows it is understood that $\omega$ is such that $n\omega$ is an integer.

Clearly for $\omega$ such that $W_{com}(\omega) \leq 0$, the convergence of $N(G, n\omega)$ to $\mathbb{E}[N(G, n\omega)]$ follows trivially. Hence in the rest of the paper our focus will be on the weight $\omega$ such that $W_{com}(\omega) > 0$.

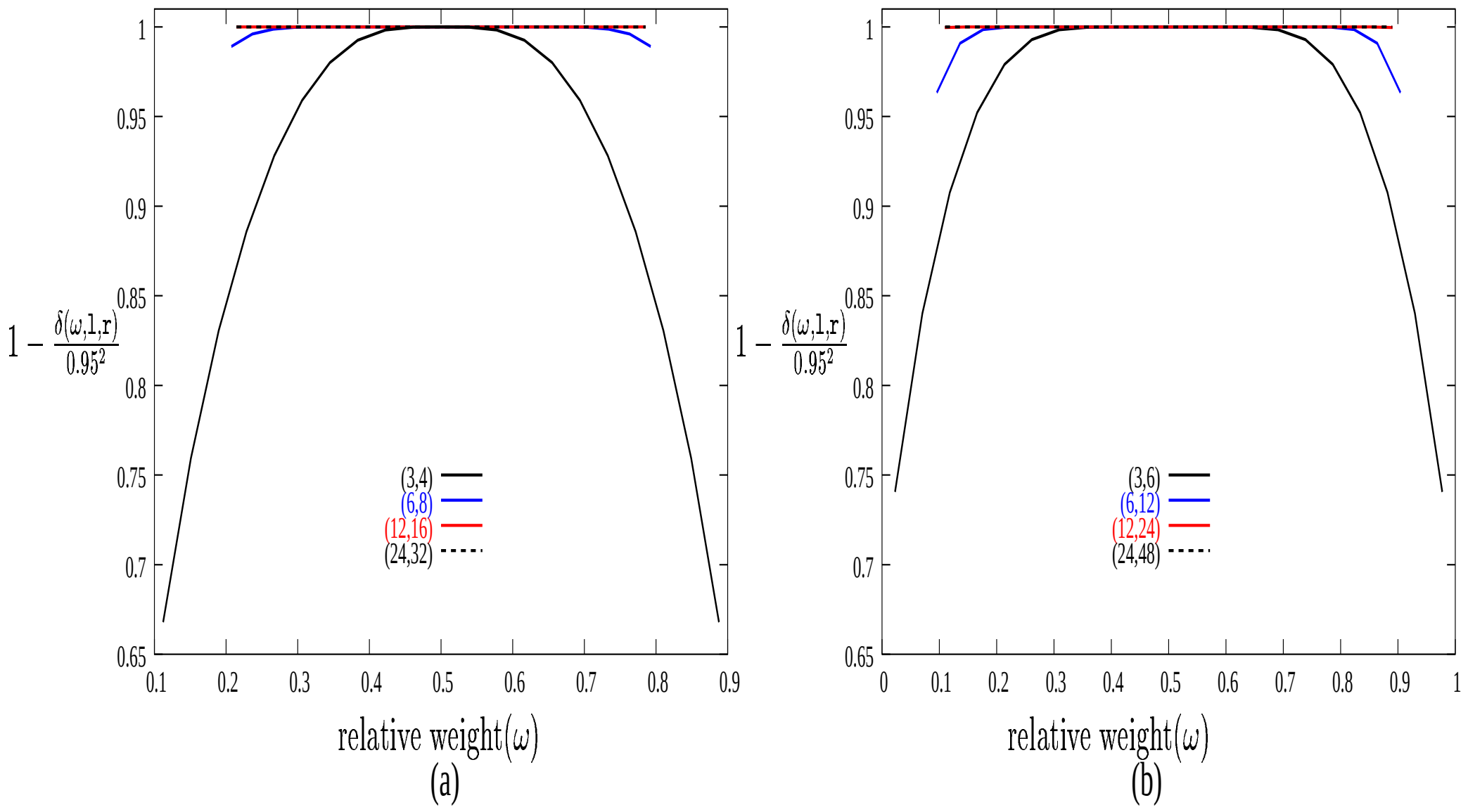

***Figure 1:*** *Plot of the bound* $1 - \frac{\delta(\omega,\mathtt{l},\mathtt{r})}{\epsilon^2}$ *for weight distribution with* $\epsilon = 0.95$, *(a) for ensembles with rate*=0.25, *(b) for ensembles with rate*=0.5.

Note that the convergence implied by the bound in (1) is pointwise. For a fixed $\omega$, (1) implies that asymptotically at least a fraction $1 - \frac{\delta(\omega,\mathtt{l},\mathtt{r})}{\epsilon^2}$ of codes in the ensemble have their weight distribution function in a window of width $\epsilon$ around the ensemble average. In Fig. 1 we plot the bound in (1) for regular codes with $\frac{\mathtt{l}}{\mathtt{r}} = 0.75$ and 0.5. We observe that if we fix the ratio $\frac{\mathtt{l}}{\mathtt{r}}$ and let $\mathtt{l}, \mathtt{r}$ increase then the bound converges to 1. This implies that for large left and right degrees, almost all the codes in the ensemble have their weight distribution very close to the ensemble average. Note that in this case it is well known that the weight distribution converges to the weight distribution of Shannon's random ensemble [12].

Another important property of LDPC codes is the stopping set distribution. Stopping sets determine the performance of LDPC codes under iterative decoding over erasure channel. The bound obtained in (1) can be easily extended to stopping set distribution. This is because of the fact that the method of determining the moments in both the cases is same. Hence we focus on the weight distribution and in the end we briefly describe the computation for stopping set distribution. In Fig. 2 we plot the bound in (1) for stopping set distribution. Again we observe that if we fix the ratio $\frac{\mathtt{l}}{\mathtt{r}}$ and let $\mathtt{l},\mathtt{r}$ increase then the bound converges to 1.

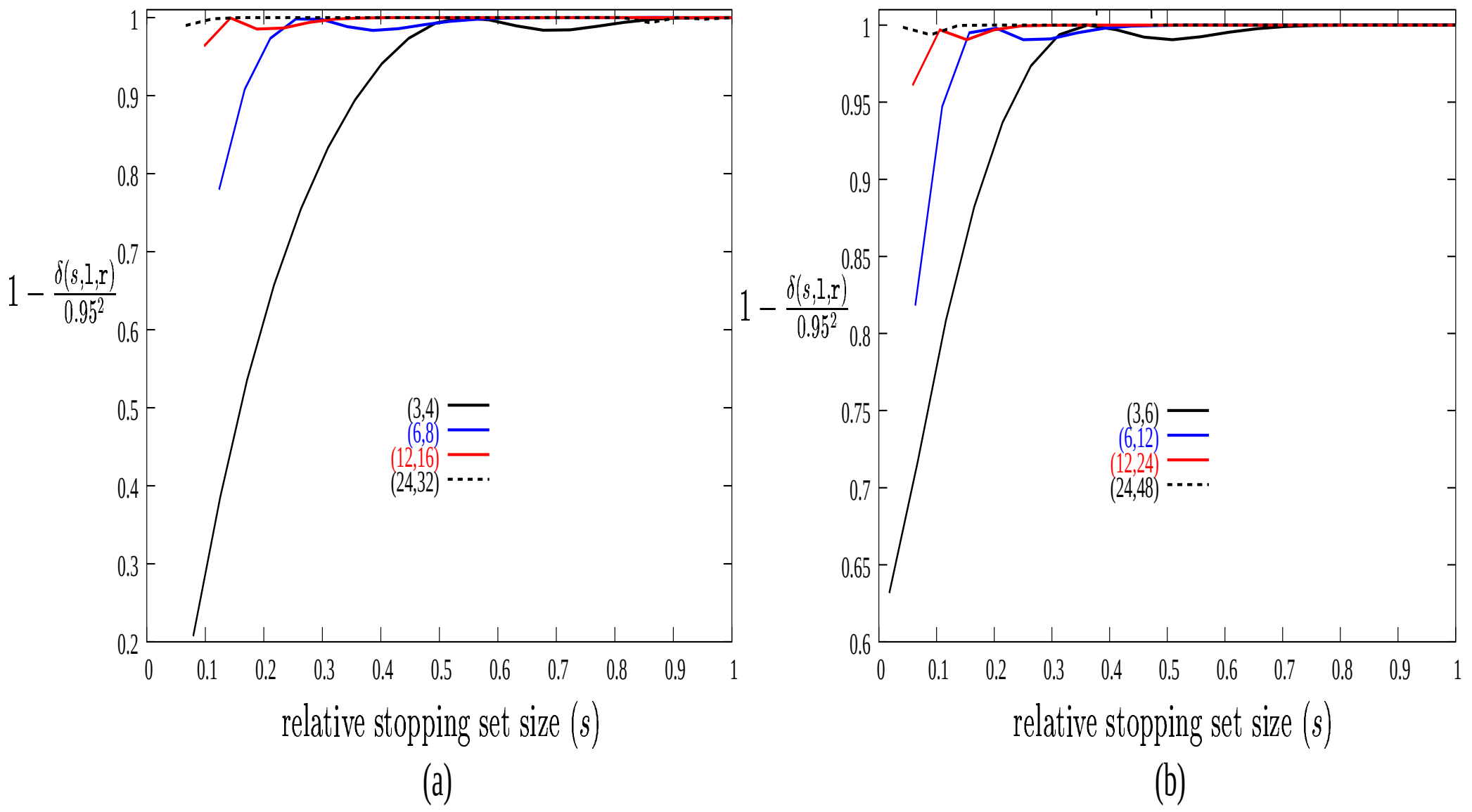

***Figure 2:*** *Plot of the bound* $1-\frac{\delta(s,\mathtt{l},\mathtt{r})}{\epsilon^2}$ *for stopping set distribution with* $\epsilon = 0.95$*, (a) for ensembles with rate*=0.25*, (b) for ensembles with rate*=0.5.

The paper is organized in the following way. A brief introduction to LDPC codes and second moment method is given in Section 2. In Section 3, we use the second moment method to prove the bound in (1) for weight distribution. We apply the second moment method to stopping set distribution in section 4. A discussion in Section 5 concludes the paper.

# 2 Preliminaries

## 2.1 LDPC Ensembles

LDPC codes, originally invented by Gallager [8], are usually defined in terms of ensembles of *bipartite graphs*. A graph consists of a set of *variable* nodes and a set of *check* nodes, together with edges connecting both sets giving rise to a code of block length $n$ in the following way: a vector $(x_1,\cdots,x_n) \in \mathrm{GF}(2)^n$ is a codeword if and only if for each check node the sum (modulo 2) of the values of its adjacent variable nodes is zero. The coordinates of a codeword are indexed by the variable nodes $1,\cdots,n$. A stopping set is a subset of the set of variable nodes such that its neighboring check nodes are connected to it at least twice.

An ensemble of bipartite graphs can be defined in terms of a pair of *degree distributions*. A degree distribution is a real valued polynomial with non-negative coefficients and it evaluates to unity at unity. Associated with the ensemble is a degree distribution pair $(\lambda(x) = \Sigma_i \lambda_i x^{i-1}, \rho(x) = \Sigma_j \rho_j x^{j-1})$, shorthand $(\lambda, \rho)$, where $\lambda_i$ $(\rho_j)$ denotes the fraction of the total number of edges connected to a variable (check) node of degree $i$ $(j)$. Given a pair $(\lambda, \rho)$ of degree distributions and the block length $n$, an *ensemble* of bipartite graphs $\mathbb{G}(n, \lambda, \rho)$ is defined by running over all possible permutations of edges connecting variable and check nodes according to $\lambda$ and $\rho$, respectively. For a $(\mathtt{l}, \mathtt{r})$-regular code ensemble $\mathbb{G}(n, \mathtt{l}, \mathtt{r})$ we have: $\lambda(x) = x^{\mathtt{l}-1}, \rho(x) = x^{\mathtt{r}-1}$. Let $G$ be a graph chosen at random from $\mathbb{G}(n, \mathtt{l}, \mathtt{r})$. Let $N(G, n\omega)$ be the weight distribution function denoting the number of codewords of weight $n\omega$ in $G$ where $\omega = \frac{W}{n}$ is the normalized weight with $W$ denoting the weight. Let $\sigma^2(G, n\omega)$ denote the variance of $N(G, n\omega)$ over the ensemble $\mathbb{G}(n, \mathtt{l}, \mathtt{r})$, $\sigma^2(G, n\omega) = \mathbb{E}[N(G, n\omega)^2] - \mathbb{E}[N(G, n\omega)]^2$. Similarly let $S(G, ns)$ denote the number of stopping sets of size $ns$ and let $\sigma^2(G, ns)$ denote the variance of $S(G, ns)$ over the ensemble $\mathbb{G}(n, \mathtt{l}, \mathtt{r})$. The *support set* of a word is the set of its non zero bits. The *overlap* between two words is the intersection of their support sets. We denote a vector $(x_1, x_2, x_3)$ by $\underline{x}$, the transpose of $\underline{x}$ by $\underline{x}^T$, the dot product between $\underline{x}$ and $\underline{y}$ is denoted by $\underline{x}.\underline{y}^T$, $\underline{x}\underline{y}$ denotes the component wise multiplication, i.e., the vector $(x_1 y_1, x_2 y_2, x_2 y_3)$. We use the notation that a vector to the power of a vector and also a scalar to the power of a vector is a vector i.e. , $\underline{x}^{\underline{k}} := (x_1^{k_1}, x_2^{k_2}, x_3^{k_3})$ and $e^{\underline{x}} := (e^{x_1}, e^{x_2}, e^{x_3})$. Finally, $x^+ := \max(x, 0)$ and $f'(t)$ denotes the derivative of the function $f(x)$ evaluated at $t$.

## 2.2 Second Moment Method

Let $\{X_n\}$ be a sequence of random variables indexed by $n$, $n \in \mathbb{N}$. Let $\sigma_n^2 = \mathbb{E}[(X_n - \mathbb{E}[X_n])^2]$ be the variance of $X_n$. Then by Chebyshev's inequality we have for any $a \geq 0$,

$$\mathrm{P}(|X_n - \mathbb{E}[X_n]| \geq a) \leq \frac{\sigma_n^2}{a^2}.$$

If we choose $a = \epsilon \mathbb{E}[X_n]$ and if $\lim_{n\to\infty} \frac{\sigma_n^2}{\mathbb{E}[X_n]^2} = \delta$, then we can draw the conclusion that

$$\lim_{n\to\infty} \mathrm{P}\left(1 - \epsilon \leq \frac{X_n}{\mathbb{E}[X_n]} \leq 1 + \epsilon\right) \geq 1 - \frac{\delta}{\epsilon^2}.$$

In order to apply this bound to $N(G, n\omega)$, we need to compute the ratio $\lim_{n\to\infty} \frac{\sigma^2(G,n\omega)}{\mathbb{E}[N(G,n\omega)]^2} = \lim_{n\to\infty} \frac{\mathbb{E}[N^2(G,n\omega)]}{\mathbb{E}[N(G,n\omega)]^2} - 1$.

# 3 Moment Calculations for Weight Distribution

We start with the first moment. As shown in [3, 5, 10],

$$\mathbb{E}[N(G, n\omega)] = \frac{\binom{n}{n\omega}}{\binom{n\mathtt{l}}{n\mathtt{l}\omega}} \mathrm{Coeff}\left(p(x)^{\frac{\mathtt{l}n}{\mathtt{r}}}, x^{n\mathtt{l}\omega}\right), \tag{2}$$

where $\mathrm{Coeff}(p(x)^{\frac{n\mathtt{l}}{\mathtt{r}}}, x^{n\mathtt{l}\omega})$ denotes the coefficient of $x^{n\mathtt{l}\omega}$ in the Taylor series expansion of $p(x)^{\frac{n\mathtt{l}}{\mathtt{r}}}$ and $p(x) = ((1+x)^{\mathtt{r}} + (1-x)^{\mathtt{r}})/2$. We note that $p(x)$ has only even powers of $x$. To remove this periodicity of powers, we define the polynomial $q(y) = p(x)$, where $y = x^2$. Now, $\mathrm{Coeff}(p(x)^{\frac{n\mathtt{l}}{\mathtt{r}}}, x^{n\mathtt{l}\omega}) = \mathrm{Coeff}(q(y)^{\frac{n\mathtt{l}}{\mathtt{r}}}, y^{\frac{n\mathtt{l}\omega}{2}})$. In the next lemma, we recall the Hayman method to approximate $\mathrm{Coeff}(q(y)^{\frac{n\mathtt{l}}{\mathtt{r}}}, y^{\frac{n\mathtt{l}\omega}{2}})$ for large values of $n$, a proof of which can be found in [6, 7].

**Lemma 3.1** [Hayman Method] Let $q(y) = \sum_i q_i y^i$ be a polynomial with non negative coefficients such that $q_0 \neq 0$ and $q_1 \neq 0$. Define $a_q(y) := y \frac{dq(y)}{dy} \frac{1}{q(y)}$ and $b_q(y) := y \frac{da_q(y)}{dy}$. Then for $n$ tending to infinity so that $\frac{n\mathtt{l}\omega}{2} \in \mathbb{N}$

$$\mathrm{Coeff}(q(y)^{\frac{n\mathtt{l}}{\mathtt{r}}}, y^{\frac{n\mathtt{l}\omega}{2}}) = \frac{q(y_\omega)^{\frac{n\mathtt{l}}{\mathtt{r}}}}{(y_\omega)^{\frac{n\mathtt{l}\omega}{2}} \sqrt{2\pi \frac{n\mathtt{l}}{\mathtt{r}} b_q(y_\omega)}} (1 + o(1)), \tag{3}$$

where the term $o(1)$ converges to zero and $y_\omega$ is the unique positive solution of $a_q(y) = \frac{\mathtt{r}\omega}{2}$.

Since $q(y) = p(x)$ and $y = x^2$, we have $a_q(y) = a_p(x)/2$, where $a_p(x) = x \frac{dp(x)}{dx} \frac{1}{p(x)}$. Similarly, $b_q(y) = b_p(x)/4$, where $b_p(x) = x \frac{da_p(x)}{dx}$. Also $y_\omega = x_\omega^2$, where $x_\omega$ is the unique positive solution of $a_p(x) = r\omega$ which simplifies to,

$$x \frac{(1+x)^{\mathtt{r}-1} - (1-x)^{\mathtt{r}-1}}{(1+x)^{\mathtt{r}} + (1-x)^{\mathtt{r}}} = \omega. \tag{4}$$

Thus by substituting these relationships in Lemma 3.1, we get

$$\mathrm{Coeff}\left(p(x)^{\frac{n\mathtt{l}}{\mathtt{r}}}, x^{n\mathtt{l}\omega}\right) = \frac{2p(x_\omega)^{\frac{n\mathtt{l}}{\mathtt{r}}}}{(x_\omega)^{n\mathtt{l}\omega} \sqrt{2\pi \frac{n\mathtt{l}}{\mathtt{r}} b_p(x_\omega)}} (1 + o(1)). \tag{5}$$

We summarize our results thus far.

**Lemma 3.2** [Ensemble Average of Weight Distribution] Consider the regular LDPC ensemble $\mathbb{G}(n, \mathtt{l}, \mathtt{r})$. Then for $\omega \in (0,1)$ such that $\mathtt{l}n\omega \in 2\mathbb{N}$,

$$\mathbb{E}[N(G, n\omega)] = \frac{2\sqrt{\mathtt{r}}}{\sqrt{2\pi n b_p(x_\omega)}} e^{n(\frac{\mathtt{l}}{\mathtt{r}} \ln(p(x_\omega)) - (\mathtt{l}-1)h(\omega) - \mathtt{l}\omega \ln(x_\omega))} (1 + o(1)),$$

where $h(\omega) = -(\omega \ln \omega + (1-\omega)\ln(1-\omega))$, $\ln \omega$ is the natural logarithm of $\omega$ and $x_\omega$ is the unique positive solution of equation (4). If $n\mathtt{l}\omega$ is odd, then $\mathbb{E}[N(G, n\omega)] = 0$.

*Proof.* We note that $n\mathtt{l}\omega$ must be even, otherwise $\mathbb{E}[N(G, n\omega)] = 0$ as $\mathrm{Coeff}\left(p(x)^{\frac{\mathtt{l}n}{\mathtt{r}}}, x^{n\mathtt{l}\omega}\right) = 0$ in (2). When $n\mathtt{l}\omega$ is even, using Stirling's approximation we get:

$$\binom{n}{n\omega} = \frac{e^{nh(\omega)}}{\sqrt{2\pi n\omega(1-\omega)}} (1 + o(1)).$$

By substituting this and (5) in (2), we get the desired result. □

To compute the second moment, we note that $\mathbb{E}[N^2(G, n\omega)] = \mathbb{E}[\sum_{w,w'} I_{w,w'}(G, n\omega)]$, where $w, w'$ are both words of length $n$ and weight $n\omega$ and

$$I_{w,w'}(G, n\omega) = \begin{cases} 1, & \text{if } w, w' \text{ are codewords of } G, \\ 0, & \text{otherwise}. \end{cases}$$

By definition of the ensemble, the expectation $\mathbb{E}[I_{w,w'}(G,n\omega)]$ does not depend on the specific choice of the pair $w,w'$ but only on the cardinality of the overlap between the support sets of $w$ and $w'$. In particular we can fix $w$ to be a codeword of weight $n\omega$ with support set $\mathcal{W} = \{1,2,\cdots,n\omega\}$, so that

$$\mathbb{E}[N^2(G,n\omega)] = \binom{n}{n\omega}\sum_{w'}\mathbb{E}[I_{w'}(G,n\omega)],$$

where we have dropped the subscript $w$ as $w$ is fixed. We can also fix $w'$ to $w'(i)$ for a given cardinality of overlap $i$ with $w$. $w'(i)$ has support set $\mathcal{W}' = \{1,2,\cdots,i,n\omega+1,\cdots,2n\omega-i\}$. Then,

$$\mathbb{E}[N^2(G,n\omega)] = \binom{n}{n\omega}\sum_{i=0}^{n\omega}\binom{n\omega}{i}\binom{n-n\omega}{n\omega-i}\mathbb{E}[I_{w'(i)}(G,n\omega)].$$

The binomials inside the summation correspond to the number of words having cardinality of overlap with $w$ equals to $i$. To calculate $\mathbb{E}[I_{w'(i)}(G,n\omega)]$, we note that there are 3 different types of edges taking value 1. These types are: edges connected to $\mathcal{W}\cap\mathcal{W}'$, edges connected to $\mathcal{W}\backslash(\mathcal{W}\cap\mathcal{W}')$ and finally, edges connected to $\mathcal{W}'\backslash(\mathcal{W}\cap\mathcal{W}')$. A placement of edges is *valid* if each check node is connected to an even number of edges from $\mathcal{W}$ as well as from $\mathcal{W}'$, i.e., if the number of edges from each of the 3 different classes are *all* even or *all* odd. A moment's thought shows that the generating function for the number of valid placement is given by $f(x_1,x_2,x_3)^{\frac{n\mathtt{l}}{\mathtt{r}}} = f(\underline{x})^{\frac{n\mathtt{l}}{\mathtt{r}}}$, where $x_1$ corresponds to the number of edges connected to $\mathcal{W}\backslash(\mathcal{W}\cap\mathcal{W}')$, $x_2$ corresponds to the number of edges connected to $\mathcal{W}\cap\mathcal{W}'$ and $x_3$ corresponds to the number of edges connected to $\mathcal{W}'\backslash(\mathcal{W}\cap\mathcal{W}')$, and where $f(\underline{x})$ is the summation of the terms in the expansion of $(1+x_1+x_2+x_3)^{\mathtt{r}}$ which have powers of $x_1,x_2$ and $x_3$ either all even or all odd. Explicitly,

$$f(\underline{x}) = \frac{1}{4}((1+x_1+x_2+x_3)^{\mathtt{r}}+(1+x_1-x_2-x_3)^{\mathtt{r}}+(1-x_1+x_2-x_3)^{\mathtt{r}}+(1-x_1-x_2+x_3)^{\mathtt{r}}). \tag{6}$$

Since there are $\mathtt{l}(n\omega-i)$ edges connected to $\mathcal{W}\backslash(\mathcal{W}\cap\mathcal{W}')$, $\mathtt{l}i$ edges connected to $\mathcal{W}\cap\mathcal{W}'$ and $\mathtt{l}(n\omega-i)$ edges connected to $\mathcal{W}'\backslash(\mathcal{W}\cap\mathcal{W}')$, we have

$$\begin{aligned}\mathbb{E}[I_{w'(i)}(G,n\omega)] &= \frac{1}{(n\mathtt{l})!}((\mathtt{l}(n\omega-i))!)^2(\mathtt{l}i)!(n\mathtt{l}-2n\mathtt{l}\omega+\mathtt{l}i)!\\ &\qquad \text{Coeff}\left(f(\underline{x})^{\frac{n\mathtt{l}}{\mathtt{r}}}, x_1^{\mathtt{l}(n\omega-i)}x_2^{\mathtt{l}i}x_3^{\mathtt{l}(n\omega-i)}\right).\end{aligned}$$

As all the edges are labeled, the factor $(n\mathtt{l})!$ corresponds to the total number of graphs in the ensemble $\mathbb{G}(n,\mathtt{l},\mathtt{r})$. The term $(\mathtt{l}(n\omega-i))!^2$ corresponds to interchanging the positions of edges connected to $\mathcal{W}\backslash(\mathcal{W}\cap\mathcal{W}')$, as well as to $\mathcal{W}'\backslash(\mathcal{W}\cap\mathcal{W}')$, $(\mathtt{l}i)!$ corresponds to interchanging the positions of edges connected to $\mathcal{W}\cap\mathcal{W}'$, and $(\mathtt{l}(n-2n\omega+i))!$ corresponds to interchanging of the positions of edges taking value 0. Hence,

$$\begin{aligned}\mathbb{E}[N^2(G,n\omega)] &= \sum_{i=0}^{n\omega}\underbrace{\frac{\binom{n}{n\omega}}{(n\mathtt{l})!}\binom{n\omega}{i}\binom{n-n\omega}{n\omega-i}((\mathtt{l}(n\omega-i))!)^2(\mathtt{l}i)!(\mathtt{l}(n-2n\omega+i))!}_{F_i}\\ &\qquad \underbrace{\text{Coeff}\left(f(\underline{x})^{n\frac{\mathtt{l}}{\mathtt{r}}}, x_1^{\mathtt{l}(n\omega-i)}x_2^{\mathtt{l}i}x_3^{\mathtt{l}(n\omega-i)}\right)}_{C_i}.\end{aligned} \tag{7}$$

Let $S_i$ be the $i^{th}$ summation term in (7), so $S_i = F_i C_i$. Note that $S_i = 0$ for $i < (2n\omega - n)^+$ as there can not exist two words of length $n$ and weight $n\omega$ such that the cardinality of their overlap is less than $(2n\omega - n)^+$. A property of the term $S_{n\omega}$ that we will need later is

$$S_{n\omega} = \mathbb{E}[N(G, n\omega)]. \tag{8}$$

This simply follows from the fact that for $i = n\omega$, the words $w$ and $w'(i)$ are identical. Now to get a closed form expression for $\mathbb{E}[N^2(G, n\omega)]$, we use Stirling's formula to approximate the factorial terms and to approximate the Coeff function we use the following multidimensional extension of Lemma 3.1 as given in Theorem 2 of [1].

**Lemma 3.3** [Multidimensional Saddle Point Method] Let $\underline{i} := (\mathtt{l}(n\omega - i), \mathtt{l}i, \mathtt{l}(n\omega - i))$, $\underline{j} := (\mathtt{l}(n\omega - j), \mathtt{l}j, \mathtt{l}(n\omega - j))$ and $0 < \lim_{n\to\infty} \frac{i}{n} < \omega$, $f(\underline{x})$ be as defined in (6) and $\underline{t} = (t_1, t_2, t_3)$ be a positive solution of $a_f(\underline{x}) = \frac{\mathtt{r}\underline{i}}{n\mathtt{l}}$, where $a_f(\underline{x}) = (\frac{x_i \partial f}{f \partial x_i})_{i=1}^3$. Then $\mathrm{Coeff}\left(f(\underline{x})^{\frac{n\mathtt{l}}{\mathtt{r}}}, \underline{x}^{\underline{i}}\right)$ can be approximated using the saddle point method for multivariate polynomials,

$$\mathrm{Coeff}\left(f(\underline{x})^{\frac{n\mathtt{l}}{\mathtt{r}}}, \underline{x}^{\underline{i}}\right) = \frac{4 f(\underline{t})^{\frac{n\mathtt{l}}{\mathtt{r}}}}{(\underline{t})^{\underline{i}} \sqrt{(2\pi \frac{n\mathtt{l}}{\mathtt{r}})^3 \left|B_f(\underline{t})\right|}}(1 + o(1)),$$

where $B_f(\underline{x})$ is a $3 \times 3$ matrix whose elements are given by $B_{f(i,j)} = x_j \frac{\partial a_i}{\partial x_j} = B_{f(j,i)}$. Also, $\mathrm{Coeff}\left(f(\underline{x})^{\frac{n\mathtt{l}}{\mathtt{r}}}, \underline{x}^{\underline{j}}\right)$ can be approximated in terms of $\mathrm{Coeff}\left(f(\underline{x})^{\frac{n\mathtt{l}}{\mathtt{r}}}, \underline{x}^{\underline{i}}\right)$. This approximation is called the *local limit theorem* of $\underline{j}$ around $\underline{i}$. Explicitly, if $\underline{u} := \sqrt{\frac{\mathtt{r}}{n\mathtt{l}}}(\underline{j} - \underline{i})$ and $\|\underline{u}\| = O((\ln n)^{\frac{1}{3}})$, then

$$\mathrm{Coeff}\left(f(\underline{x})^{\frac{n\mathtt{l}}{\mathtt{r}}}, \underline{x}^{\underline{j}}\right) = \underline{t}^{\underline{i}-\underline{j}} \mathrm{Coeff}\left(f(\underline{x})^{\frac{n\mathtt{l}}{\mathtt{r}}}, \underline{x}^{\underline{i}}\right) \exp\left(-\frac{1}{2}\underline{u}.B_f(\underline{t})^{-1}.\underline{u}^T\right)(1 + o(1)).$$

*Proof.* The proof of the lemma follows from a modification of the proof of Theorem 2 of [1]. This is rather tedious and is therefore relegated to the appendix. □

The system of equations corresponding to $a(\underline{x}) = \frac{\mathtt{r}\underline{i}}{n\mathtt{l}}$ is symmetric in $x_1$ and $x_3$. Hence a positive solution $\underline{x}$ of this system of equations satisfies $x_1 = x_3$ and the system reduces to the following equations,

$$x_1 \frac{(1 + 2x_1 + x_2)^{\mathtt{r}-1} - (1 - 2x_1 + x_2)^{\mathtt{r}-1}}{(1 + 2x_1 + x_2)^{\mathtt{r}} + 2(1 - x_2)^{\mathtt{r}} + (1 - 2x_1 + x_2)^{\mathtt{r}}} = \omega - \alpha, \tag{9}$$

$$x_2 \frac{(1 + 2x_1 + x_2)^{\mathtt{r}-1} - 2(1 - x_2)^{\mathtt{r}-1} + (1 - 2x_1 + x_2)^{\mathtt{r}-1}}{(1 + 2x_1 + x_2)^{\mathtt{r}} + 2(1 - x_2)^{\mathtt{r}} + (1 - 2x_1 + x_2)^{\mathtt{r}}} = \alpha, \tag{10}$$

where $\alpha = \frac{i}{n}$.

In order to evaluate the second moment, we need to find the dominant terms of the summation in (7). To find all the dominant terms, let the term corresponding to $i = i_m$ i.e. $S_{i_m} = F_{i_m} C_{i_m}$ be a local maximum of $\{S_i\}_{i=0}^{n\omega}$. We first check if the end terms $S_{(2n\omega - n)^+}$ and $S_{n\omega}$ can be dominant. The assumption 2 of the Lemma 3.4 eliminates the possibility that $S_{(2n\omega-n)^+}$ is a dominant term. In the proof of Lemma 3.4 we will see that $\ln(S_{n\omega^2})/n = 2W_{\mathrm{com}}(\omega)$. This with eqn(8) implies that $S_{n\omega}$ is not a dominant term. So we consider $i_m$ such that $0 < \lim_{n\to\infty} \frac{i_m}{n} < \omega$. Let $\Delta = i - i_m$ and $\alpha_m = \frac{i_m}{n}$. We expand $F_i$ and $C_i$ for $\Delta \in (-\sqrt{n}(\ln n)^{\frac{1}{3}}, \sqrt{n}(\ln n)^{\frac{1}{3}})$ in terms of $F_{i_m}$ and $C_{i_m}$ using Stirling's approximation and the local limit theorem of Lemma 3.3 respectively. Then,

$$
\begin{aligned}
F_i &= F_{i_m} \exp\left(\Delta(\mathtt{l}-1)\ln\left(\frac{i_m(n-2n\omega+i_m)}{(n\omega-i_m)^2}\right)\right) \\
&\quad \exp\left(\Delta^2\left(\frac{\mathtt{l}-1}{n\omega-i_m}+\frac{\mathtt{l}-1}{2i_m}+\frac{\mathtt{l}-1}{2(n-2n\omega+i_m)}\right)\right)\left(1+O\left(\frac{\Delta^3}{n^2}\right)\right), \\
C_i &= C_{i_m}\exp\left(\Delta\mathtt{l}\ln\left(\frac{t_1^2}{t_2}\right)-\frac{\Delta^2}{2n\sigma_c^2(\alpha_m)}\right)(1+o(1)),
\end{aligned}
$$

where

$$
\begin{aligned}
F_{i_m} &= \mathtt{l}\sqrt{\mathtt{l}}\left(\frac{(n\omega-i_m)^{2(n\omega-i_m)}i_m^{i_m}(n-2n\omega+i_m)^{n-2n\omega+i_m}}{n^n}\right)^{\mathtt{l}-1}(1+o(1)), \\
\sigma_c^2(\alpha_m) &= \frac{1}{\mathtt{lr}\left((-1,1,-1).B_f(\underline{t})^{-1}.(-1,1,-1)^T\right)}, \qquad (11)
\end{aligned}
$$

Hence,

$$
\begin{aligned}
S_i &= \mathrm{S}_{i_m}\exp\left(\Delta\left((\mathtt{l}-1)\ln\left(\frac{i_m(n-2n\omega+i_m)}{(n\omega-i_m)^2}\right)+\mathtt{l}\ln\left(\frac{t_1^2}{t_2}\right)\right)\right) \\
&\quad .\exp\left(\Delta^2\left(\frac{\mathtt{l}-1}{n\omega-i_m}+\frac{\mathtt{l}-1}{2i_m}+\frac{\mathtt{l}-1}{2(n-2n\omega+i_m)}-\frac{1}{2n\sigma_c^2(\alpha_m)}\right)\right)(1+o(1)). \qquad (12)
\end{aligned}
$$

We know that there is a local maximum at $\Delta = 0$, hence the coefficient of $\Delta$ in (12) will vanish. This gives an additional equation governing $\alpha_m$:

$$
\left(\frac{\alpha_m(1-2\omega+\alpha_m)}{(\omega-\alpha_m)^2}\right)^{\mathtt{l}-1} = \left(\frac{t_2}{t_1^2}\right)^{\mathtt{l}}. \qquad (13)
$$

We solve (9), (10) and (13) and find all the solutions such that $0 < \alpha_m < \omega$, $t_1 > 0$, $t_2 > 0$ and the coefficient of $\Delta^2$ in (12) is negative (this ensures that $S_{i_m}$ is a local maximum). One of the possible solution to this system of polynomial equations is $\alpha_m = \omega^2$. This is because $\{C_i\}_{i=0}^{n\omega}$ and $\{F_i\}_{i=0}^{n\omega}$ are concave and convex sequences respectively, both achieving their extreme values at $i = n\omega^2$. Hence $\{S_i\}_{i=0}^{n\omega}$ also achieves an extreme value at $i = n\omega^2$. If $\alpha_m = \omega^2$ is a unique global maximum in the solution set of (9), (10) and (13), then we can get a closed form expression for second moment. We summarize this in the following lemma.

**Lemma 3.4** [Second Moment Method] Consider the regular LDPC ensemble $\mathbb{G}(n,\mathtt{l},\mathtt{r})$. Then for $\omega \in (0,1)$, if $W_{\text{com}}(\omega) > 0$ and if the following conditions are satisfied,

1. $\alpha_m = \omega^2$ is the only solution of (9), (10) and (13) for which coefficient of $\Delta^2$ in (12) is negative.
2. $\lim_{n\to\infty}\frac{\ln(S_{n\omega^2})}{n} > \frac{\ln(S_{(2n\omega-n)^+})}{n}$,

then by the second moment method we have,

$$
\lim_{n\to\infty} \mathrm{P}\left(1-\epsilon \le \frac{N(G,n\omega)}{\mathbb{E}[N(G,n\omega)]} \le 1+\epsilon\right) \ge 1-\frac{\delta(\omega,\mathtt{l},\mathtt{r})}{\epsilon^2},
$$

where

$$\begin{aligned}
\delta(\omega,\mathtt{l},\mathtt{r}) &= \frac{b_p(x_\omega)\sqrt{r}\omega(1-\omega)\sigma_c(\omega^2)}{\sqrt{\left|B_f(x_\omega,x_\omega^2,x_\omega)\right|(\omega^2(1-\omega)^2-(\mathtt{l}-1)\sigma_c^2(\omega^2))}}-1,\\
\sigma_c^2(\omega^2) &= \frac{1}{\mathtt{lr}((-1,1,-1).B_f(x_\omega,x_\omega^2,x_\omega)^{-1}.(-1,1,-1)^T)},
\end{aligned}$$

and $x_\omega$ is the only positive solution of (4).

Remark: Note that the conditions of Lemma 3.4 are hard to verify in general but they are typically easy to verify for any given regular LDPC ensemble.

*Proof.* We observe that the solution $\underline{t}$ of (9), (10) for $\alpha=\omega^2$ satisfies $t_2=t_1^2$ and this system of equations reduces to a single equation which is identical to (4), the equation we need to solve to find $\mathbb{E}[N(G,n\omega)]$. Thus $t_1=x_\omega$. By (12) and noting that the terms $S_{n\omega^2+\Delta}$ for $\Delta\notin(-\sqrt{n}(\ln n)^{\frac{1}{3}},\sqrt{n}(\ln n)^{\frac{1}{3}})$ are much smaller than $S_{n\omega^2}$, we get

$$\begin{aligned}
\mathbb{E}[N^2(G,n\omega)] &= S_{n\omega^2}\sum_{\Delta=-\sqrt{n}(\ln n)^{\frac{1}{3}}}^{\sqrt{n}(\ln n)^{\frac{1}{3}}}\exp\left(\frac{-\Delta^2}{2\sigma_s^2}\right)(1+o(1)),\\
&= S_{n\omega^2}\int_{-\infty}^{\infty}\exp\left(\frac{-x^2}{2\sigma_s^2}\right)dx(1+o(1)),\\
&= S_{n\omega^2}\sqrt{2\pi\sigma_s^2}(1+o(1)),\\
\text{where }\ \frac{1}{\sigma_s^2} &= \frac{1}{n\sigma_c^2(\omega^2)}-\frac{\mathtt{l}-1}{n\omega^2(1-\omega)^2}.
\end{aligned}$$

Also $f(x_\omega,x_\omega^2,x_\omega)=p(x_\omega)^2$. To evaluate $S_{n\omega^2}$, we use Lemma 3.3 and Stirling's approximation for factorial terms. This gives,

$$\mathbb{E}[N^2(G,n\omega)]=\frac{4\sigma_c(\omega^2)\mathtt{r}\sqrt{\mathtt{r}}\omega(1-\omega)e^{2n(\frac{\mathtt{l}}{\mathtt{r}}\ln(p(x_\omega))-(\mathtt{l}-1)h(\omega)-\mathtt{l}\omega\ln(x_\omega))}}{2\pi n\sqrt{(\omega^2(1-\omega)^2-(\mathtt{l}-1)\sigma_c^2(\omega^2))\left|B_f(x_\omega,x_\omega^2,x_\omega)\right|}}(1+o(1)).$$

We need the condition $W_{\text{com}}(\omega)>0$, as $\lim_{n\to\infty}\frac{\ln(S_{n\omega^2})}{n}=2W_{\text{com}}(\omega)$ and $\lim_{n\to\infty}\frac{\ln(S_{n\omega})}{n}=W_{\text{com}}(\omega)$. Clearly when $W_{\text{com}}(\omega)$ is negative, $S_{n\omega^2}$ can not be a global maximum. Now using Lemma 3.2 the second moment method gives us:

$$\lim_{n\to\infty}\mathrm{P}\left(1-\epsilon\le\frac{N(G,n\omega)}{\mathbb{E}[N(G,n\omega)]}\le1+\epsilon\right)\quad\ge\quad1-\frac{\delta(\omega,\mathtt{l},\mathtt{r})}{\epsilon^2}.$$

This proves the lemma. □

The bound obtained in Lemma 3.4 can in general only be evaluated numerically except for the cases when (4) can be solved analytically. For example for the $(3,4)$-regular code we get,

$$\delta(\omega,3,4)=\frac{8\omega(1-\omega)(3-\Omega)}{\sqrt{(-21+80\omega(1-\omega)+9\Omega)(81-27\Omega+16\omega(1-\omega)(8\omega-8\omega^2-18+3\Omega))}},$$

where $\Omega=\sqrt{9-32\omega+32\omega^2}$.

# 4 Moment Calculations for Stopping Set Distribution

As shown in [14], the first moment of stopping set distribution is given by

$$\mathbb{E}[S(G,ns)] = \frac{\binom{n}{ns}}{\binom{n\mathtt{l}}{n\mathtt{l}s}} \text{Coeff}\left((\beta(x))^{\frac{\mathtt{l}n}{\mathtt{r}}}, x^{n\mathtt{l}s}\right),$$

where $\beta(x) = (1+x)^{\mathtt{r}} - \mathtt{r}x$. Applying the Lemma 3.1 and using Stirling's approximation we get

$$\mathbb{E}[S(G,ns)] = \frac{\sqrt{\mathtt{r}}}{\sqrt{2\pi n b_{\beta}(x_s)}} e^{n\left(\frac{1}{\mathtt{r}}\ln(\beta(x_s)) - (\mathtt{l}-1)h(s) - \mathtt{l}s\ln(x_s)\right)}(1+o(1)),$$

where $x_s$ is the only positive solution of

$$x\frac{(1+x)^{\mathtt{r}-1}-1}{(1+x)^{\mathtt{r}} - \mathtt{r}x} = s. \tag{14}$$

The second moment is $\mathbb{E}[S^2(G,ns)] = \mathbb{E}[\sum_{\mathtt{s},\mathtt{s}'} I_{\mathtt{s},\mathtt{s}'}(G,ns)]$, where $\mathtt{s},\mathtt{s}'$ are both stopping sets of cardinality $ns$. By definition of the ensemble, the expectation $\mathbb{E}[I_{\mathtt{s},\mathtt{s}'}(G,ns)]$ depends only on the cardinality of the overlap between $\mathtt{s}$ and $\mathtt{s}'$. Hence like in the previous section for weight distribution we can fix $\mathtt{s}$ to be equal to $\mathtt{s} = \{1,2,\cdots,ns\}$ and for a given cardinality of overlap $i$ we can fix $\mathtt{s}'$ to be equal to $\mathtt{s}'(i) = \{1,2,\cdots,i,ns+1,\cdots,2ns-i\}$. As $\mathtt{s}$ is fixed, we drop the subscript $\mathtt{s}$ in $I_{\mathtt{s},\mathtt{s}'}(G,ns)$. This gives

$$\mathbb{E}[S^2(G,ns)] = \binom{n}{ns}\sum_{i=0}^{ns}\binom{ns}{i}\binom{n-ns}{ns-i}\mathbb{E}[I_{\mathtt{s}'(i)}(G,ns)].$$

For $I_{\mathtt{s}'(i)}(G,ns) = 1$ we need that every check node in $G$ is either not connected to $\mathtt{s}$ or connected to $\mathtt{s}$ by more than one edge. Similarly every check node is either not connected to $\mathtt{s}'(i)$ or connected to $\mathtt{s}'(i)$ by more than one edge. This implies

$$\mathbb{E}[I_{\mathtt{s}'(i)}(G,ns)] = \frac{1}{(n\mathtt{l})!}((\mathtt{l}(ns-i))!)^2(\mathtt{l}i)!(n\mathtt{l}-2n\mathtt{l}s+\mathtt{l}i)!\text{Coeff}\left(g(\underline{x})^{\frac{n\mathtt{l}}{\mathtt{r}}}, x_1^{\mathtt{l}(ns-i)}x_2^{\mathtt{l}i}x_3^{\mathtt{l}(ns-i)}\right),$$

where

$$\begin{aligned} g(\underline{x}) \quad = \quad & (1+x_1+x_2+x_3)^{\mathtt{r}} - \mathtt{r}(1+x_1)^{\mathtt{r}-1}(x_2+x_3) - \mathtt{r}x_1\left((1+x_3)^{\mathtt{r}-1} - (\mathtt{r}-1)x_3\right) \\ & -\mathtt{r}x_2\left((1+x_3)^{\mathtt{r}-1} - 1\right). \end{aligned} \tag{15}$$

Thus

$$\begin{aligned} \mathbb{E}[S^2(G,ns)] \quad = \quad & \sum_{i=0}^{ns}\underbrace{\frac{\binom{n}{ns}}{(n\mathtt{l})!}\binom{ns}{i}\binom{n-ns}{ns-i}((\mathtt{l}(ns-i))!)^2(\mathtt{l}i)!(\mathtt{l}(n-2ns+i))!}_{F_i} \\ & \underbrace{\text{Coeff}\left(g(\underline{x})^{n\frac{\mathtt{l}}{\mathtt{r}}}, x_1^{\mathtt{l}(ns-i)}x_2^{\mathtt{l}i}x_3^{\mathtt{l}(ns-i)}\right)}_{C_i}. \end{aligned} \tag{16}$$

To evaluate Coeff in (16) we use Theorem 2 of [1]. Again applying the same line of arguments as for the weight distribution and if the conditions of Lemma 3.4 are true in the setting of stopping set distribution, then we get

$$\mathbb{E}[S^2(G,ns)] = \frac{\sigma_c(s^2)\mathtt{r}\sqrt{\mathtt{r}}s(1-s)e^{2n(\frac{1}{\mathtt{r}}\ln(\beta(x_s))-(\mathtt{l}-1)h(s)-\mathtt{l}s\ln(x_s))}}{2\pi n\sqrt{(s^2(1-s)^2-(\mathtt{l}-1)\sigma_c^2(s^2))\,|B_g(x_s,x_s^2,x_s)|}}(1+o(1)).$$

where $B_g(\underline{x})$ is same as defined in Lemma 3.3 with respect to $g(\underline{x})$ and $x_s$ is the positive solution of (14). Hence by the second moment method we have,

$$\lim_{n\to\infty} \mathrm{P}\left(1-\epsilon \le \frac{S(G,ns)}{\mathbb{E}[N(G,ns)]} \le 1+\epsilon\right) \ge 1 - \frac{\delta(s,\mathtt{l},\mathtt{r})}{\epsilon^2},$$

where

$$\delta(s,\mathtt{l},\mathtt{r}) = \frac{b_\beta(x_s)\sqrt{r}s(1-s)\sigma_c(s^2)}{\sqrt{|B_g(x_s,x_s^2,x_s)|\,(s^2(1-s)^2-(\mathtt{l}-1)\sigma_c^2(s^2))}} - 1,$$

$$\sigma_c^2(s^2) = \frac{1}{\mathtt{lr}\,((-1,1,-1).B_g(x_s,x_s^2,x_s)^{-1}.(-1,1,-1)^T)}.$$

# 5 Discussion

Fix the relative weight $\omega$. If $\epsilon \in (0,1)$ then we conclude that asymptotically for at least a fraction $1-\frac{\delta(\omega,\mathtt{l},\mathtt{r})}{\epsilon^2}$ of codes, the number of codewords $N(G,n\omega)$ (for a fixed $\omega$) is at most a constant factor away from the ensemble average.

| $(\mathtt{l},\mathtt{r})$-code | $\omega_{min}$ | $1-\frac{\delta(\omega_{min},\mathtt{l},\mathtt{r})}{0.95^2}$ |
|---|---|---|
| $(3,6)$ | 0.0227334 | 0.740611 |
| $(6,12)$ | 0.0956337 | 0.963306 |
| $(12,24)$ | 0.109404 | 0.999617 |
| $(24,48)$ | 0.110026 | $\sim 1.0$ |

***Table 1:*** $\lim_{\omega\to\omega_{min}+} 1-\frac{\delta(\omega,\mathtt{l},\mathtt{r})}{\epsilon^2}$ *for rate* $=\frac{1}{2}$ *and* $\epsilon = 0.95$.

| $(\mathtt{l},\mathtt{r})$-code | $\omega_{min}$ | $1-\frac{\delta(\omega_{min},\mathtt{l},\mathtt{r})}{0.95^2}$ |
|---|---|---|
| $(3,4)$ | 0.112159 | 0.667889 |
| $(6,8)$ | 0.207437 | 0.989098 |
| $(12,16)$ | 0.214428 | 0.999994 |
| $(24,32)$ | 0.214502 | $\sim 1.0$ |

***Table 2:*** $\lim_{\omega\to\omega_{min}+} 1-\frac{\delta(\omega,\mathtt{l},\mathtt{r})}{\epsilon^2}$ *for rate* $=\frac{1}{4}$ *and* $\epsilon = 0.95$.

Also from Fig. 1 we see that $1-\frac{\delta(\omega,\mathtt{l},\mathtt{r})}{\epsilon^2}$ is an increasing function of $\omega$ for $\omega \in (\omega_{min}, 0.5)$ and is a decreasing function for $\omega > 0.5$. It is equal to 1 for $\omega = 0.5$. This implies that asymptotically in almost all the codes there are $\mathbb{E}[N\left(G,\frac{n}{2}\right)](1\pm\epsilon)$ codewords of weight $\frac{n}{2}$. For $\omega$ close to the typical minimum distance $\omega_{min}$, the bound stays nontrivial. In Table 1 and 2, $\lim_{\omega\to\omega_{min}+} 1 - \frac{\delta(\omega_{min},\mathtt{l},\mathtt{r})}{0.95^2}$ is given for regular codes of rate 0.5 and 0.25 respectively. We observe that if we fix the rate and let $\mathtt{l}$ and $\mathtt{r}$ increase then the bound approaches 1 for all $\omega$ for which $W_{\text{com}}(\omega)$ is positive. This implies that for regular ensembles with large left and right degree almost all the codes have a weight distribution which is very close to the ensemble average. We observe the same phenomenon for stopping set distribution. We see that the second moment method can capture the concentration property of the weight distribution and stopping set distribution for regular ensembles with large left and right degrees. However for the regular ensembles in general it fails to do so. Potentially by applying more sophisticated methods one could obtain better bounds, e.g., the second moment method with conditioning [9] or other methods given in [11].

**Acknowledgment** The author wishes to thank Rüdiger Urbanke for many helpful discussions and suggestions on the preparation of the paper. The author would also like to thank anonymous reviewers for their comments which helped in improving the quality of the paper.

# 6 Appendix

We modify the proof of Theorem 2 of [1] to prove Lemma 3.3. Let $\varphi_n(\underline{z}) = f(\underline{z})^{\frac{n1}{r}}$, $R = [-\pi,\pi]^3$ and $I = \sqrt{-1}$. We also expand $\varphi_n(\underline{z})$ as $\varphi_n(\underline{z}) = \sum_{\underline{k}} a_n(\underline{k})\underline{z}^{\underline{k}}$. Let $\underline{t}$ be the positive solution of $a_f(\underline{x}) = \frac{ri}{n1}$. From the inverse Fourier transform, we get

$$\frac{1}{(2\pi)^3}\int_R \frac{\varphi_n(\underline{t}e^{I\underline{v}})}{\varphi_n(\underline{t})} e^{-I\underline{j}.\underline{v}^T}\, d\underline{v} = \frac{a_n(\underline{j})\underline{t}^{\underline{j}}}{\varphi_n(\underline{t})}. \tag{17}$$

We recall that the Fourier transform of a Gaussian is again a Gaussian,

$$\int_{-\infty}^{\infty}\int_{-\infty}^{\infty}\int_{-\infty}^{\infty} e^{-I\underline{u}.\underline{s}^T - \frac{\underline{s}.B_f(\underline{t}).\underline{s}^T}{2}}\, d\underline{s} = \sqrt{\frac{(2\pi)^3}{|B_f(\underline{t})|}}\, e^{-\frac{1}{2}\underline{u}.B_f(\underline{t})^{-1}.\underline{u}^T}. \tag{18}$$

Also from the proof of Theorem 2 of [1], for any function $K(n)$ growing with $n$, we have

$$\left|\int_{-K(n)}^{K(n)}\int_{-K(n)}^{K(n)}\int_{-K(n)}^{K(n)} e^{-I\underline{u}.\underline{s}^T - \frac{\underline{s}.B_f(\underline{t}).\underline{s}^T}{2}}\, d\underline{s} - \int_{-\infty}^{\infty}\int_{-\infty}^{\infty}\int_{-\infty}^{\infty} e^{-I\underline{u}.\underline{s}^T - \frac{\underline{s}.B_f(\underline{t}).\underline{s}^T}{2}}\, d\underline{s}\right| = O\left(\frac{1}{K(n)}\right). \tag{19}$$

We would like to show that for $n \to \infty$

$$\left|\left(\frac{n1}{r}\right)^{\frac{3}{2}} \int_R \frac{\varphi_n(\underline{t}e^{I\underline{v}})}{\varphi_n(\underline{t})} e^{-I\underline{j}.\underline{v}^T}\, d\underline{v} - 4\sqrt{\frac{(2\pi)^3}{|B_f(\underline{t})|}}\, e^{-\frac{1}{2}\underline{u}.B_f(\underline{t})^{-1}.\underline{u}^T}\right| = o(1). \tag{20}$$

To prove this, we write $\varphi_n(\underline{t}e^{I\underline{v}})$ in exponential-log form and take the Taylor series expansion of the exponent around $\underline{v} = 0$,

$$\varphi_n(\underline{t}e^{I\underline{v}}) = \mathrm{Exp}\left(\frac{n1}{r}\left(\ln(f(\underline{t})) + Ia_f(\underline{t}).\underline{v}^T - \frac{\underline{v}.B_f(\underline{t}).\underline{v}^T}{2} + O(\|\underline{v}\|^3)\right)\right). \tag{21}$$

Note that as $\ln(\varphi_n(\underline{t}))$ is analytic, so all the third order partial derivative of $\varphi_n\left(\underline{t}e^{I\underline{v}}\right)$ are bounded. Now we partition $R$ as $R=\cup_{i=1}^{5}R_i$, where $R_1=[-\delta,\delta]^3, R_2=[-\delta,\delta]\times[\pi-\delta,\pi+\delta]^2, R_3=[\pi-\delta,\pi+\delta]\times[-\delta,\delta]\times[\pi-\delta,\pi+\delta], R_4=[\pi-\delta,\pi+\delta]^2\times[-\delta,\delta], R_5=R\backslash(R_1\cup R_2\cup R_3\cup R_4)$. Here $\delta$ can be any decaying function of $n$ which satisfies that as $n\to\infty$ then $n\delta^2\to\infty$ and $n\delta^3\to 0$. We choose $\delta=n^{-\frac{2}{5}}$. This ensures that the term $O(||\underline{v}||^3)$ is negligible. By the symmetry of $f(\underline{x})$, $\varphi_n(x_1,x_2,x_3)=\varphi_n(x_1,-x_2,-x_3)=\varphi_n(-x_1,x_2,-x_3)=\varphi_n(-x_1,-x_2,x_3)$. This implies,

$$\int_{R_1}\frac{\varphi_n(\underline{t}e^{I\underline{v}})}{\varphi_n(\underline{t})}e^{-I\underline{j}.\underline{v}^T}d\underline{v}=\int_{R_k}\frac{\varphi_n(\underline{t}e^{I\underline{v}})}{\varphi_n(\underline{t})}e^{-I\underline{j}.\underline{v}^T}d\underline{v},\qquad\text{where } k\in\{2,3,4\}.$$

Now,

$$\begin{aligned}\int_{R_1}\frac{\varphi_n(\underline{t}e^{I\underline{v}})}{\varphi_n(\underline{t})}e^{-I\underline{j}.\underline{v}^T}d\underline{v} &\overset{\substack{a_f(\underline{x})=\frac{\mathtt{r}\underline{i}}{n\mathtt{l}}\\(21)}}{=} \int_{R_1}e^{I(\underline{i}-\underline{j}).\underline{v}^T-\frac{ln}{2r}\underline{v}.B_f(\underline{t}).\underline{v}^T+O(n\delta^3)}d\underline{v},\\ &\overset{\underline{y}:=\sqrt{\frac{n\mathtt{l}}{\mathtt{r}}}\underline{v}}{=} \left(\frac{\mathtt{r}}{n\mathtt{l}}\right)^{\frac{3}{2}}\int_{R_1'}e^{-I\underline{u}.\underline{y}^T-\frac{\underline{y}.B_f(\underline{t}).\underline{y}^T}{2}}(1+O(n^{-\frac{1}{5}}))d\underline{y},\\ &\overset{(18,19)}{=} \left(\frac{\mathtt{r}}{n\mathtt{l}}\right)^{\frac{3}{2}}\sqrt{\frac{(2\pi)^3}{|B_f(\underline{t})|}}e^{-\frac{\underline{u}.B_f(\underline{t})^{-1}.\underline{u}^T}{2}}(1+O(n^{-\frac{1}{5}}))+\left(\frac{\mathtt{r}}{n\mathtt{l}}\right)^{\frac{3}{2}}O(n^{-\frac{1}{10}}).\end{aligned}$$

where $R_1'=[-\sqrt{\frac{1}{\mathtt{r}}}n^{\frac{1}{10}},\sqrt{\frac{1}{\mathtt{r}}}n^{\frac{1}{10}}]^3$. Recall that,

$$\int_{R_5}\left|\frac{\varphi_n(\underline{t}e^{I\underline{v}})}{\varphi_n(\underline{t})}e^{-I\underline{j}.\underline{v}^T}\right|d\underline{v}=\int_{R_5}\left|\frac{f(\underline{t}e^{I\underline{v}})}{f(\underline{t})}\right|^{\frac{n\mathtt{l}}{\mathtt{r}}}d\underline{v}.$$

Further $f(\underline{t})$ is a 3-variable polynomial of finite degree. Let $f(\underline{t})=\sum_{\underline{k}}b(\underline{k})\underline{t}^{\underline{k}}$. Then by some algebraic manipulation we get,

$$\left|\frac{f(\underline{t}e^{I\underline{v}})}{f(\underline{t})}\right|^2=1-\frac{\sum_{\underline{k}\neq\underline{l}}b(\underline{k})b(\underline{l})\underline{t}^{\underline{k}+\underline{l}}(1-\cos\left((\underline{k}-\underline{l}).v^T\right)}{f(\underline{t})^2}. \tag{22}$$

Also $f(\underline{t})$ has $1,t_1^2,t_2^2,t_3^2$ as its summation terms and in $R_5$ at least one of the variable $v_k$ satisfies $v_k\notin[-\delta,\delta]$ where $k\in\{1,2,3\}$. This with eqn(22) implies that for some positive constants $c,c_1$,

$$\begin{aligned}\int_{R_4}\left|\frac{f(\underline{t}e^{I\underline{v}})}{f(\underline{t})}\right|^{\frac{n\mathtt{l}}{\mathtt{r}}}d\underline{v}\le\pi^3(1-c_1\delta^2)^{\frac{n\mathtt{l}}{2\mathtt{r}}} &= \pi^3(1-c_1n^{-\frac{4}{5}})^{\frac{n\mathtt{l}}{2\mathtt{r}}},\\ &= O(e^{-cn^{\frac{1}{5}}}).\end{aligned}$$

By combining the above steps we get,

$$\begin{aligned}\left|\left(\frac{n\mathtt{l}}{\mathtt{r}}\right)^{\frac{3}{2}}\int_R\frac{\varphi_n(\underline{t}e^{I\underline{v}})}{\varphi_n(\underline{t})}e^{-I\underline{j}.\underline{v}^T}d\underline{v}-4\sqrt{\frac{(2\pi)^3}{|B_f(\underline{t})|}}e^{-\frac{1}{2}\underline{u}.B_f(\underline{t})^{-1}.\underline{u}^T}\right| &= O(n^{-\frac{1}{10}}),\\ \left|\left(\frac{n\mathtt{l}}{\mathtt{r}}\right)^{\frac{3}{2}}\frac{(2\pi)^3a_n(\underline{j})\underline{t}^{\underline{j}}}{\varphi_n(\underline{t})}-4\sqrt{\frac{(2\pi)^3}{|B_f(\underline{t})|}}e^{-\frac{1}{2}\underline{u}.B_f(\underline{t})^{-1}.\underline{u}^T}\right| &\overset{(17)}{=} O(n^{-\frac{1}{10}}).\end{aligned} \tag{23}$$

The approximation of Coeff($f(\underline{x})^{\frac{n\mathtt{l}}{\mathtt{r}}},\underline{x}^{\underline{i}}$) is obtained by substituting $\underline{j}=\underline{i}$ (which implies $\underline{u}=(0,0,0)$) in (23). Also for the local limit theorem to hold we need in (23) that $e^{\frac{1}{2}\underline{u}.B_f(\underline{t})^{-1}.\underline{u}^T}n^{-\frac{1}{10}}=o(1)$. For our application choosing $||\underline{u}||=O((\ln n)^{\frac{1}{3}})$ suffices.